\documentclass{ws-p8-50x6-00}
\usepackage{graphicx}
\def\err#1#2{\lower2pt\hbox{ $\stackrel{\scriptstyle +#1}{\scriptstyle -#2}$}}
\newcommand{\alt}{\mathrel{\raisebox{-.6ex}{$\stackrel{\textstyle<}{\sim}$}}}

\def\overlay#1#2{\ifmmode%
\setbox0=\hbox{$#1$}%
\setbox1=\hbox to\wd0{\hss$#2$\hss}\else%
\setbox0=\hbox{#1}%
\setbox1=\hbox to\wd0{\hss#2\hss}\fi%
#1\hskip-\wd0\box1 }

\begin{document}

\title{Embed Zee Neutrino Mass Model into SUSY}

\author{Kingman Cheung
\footnote{
Invited talk at the Particle Physics Phenomenology Workshop, Taitung,
Taiwan, November 2000.}
}

\address{National Center for Theoretical Science, National Tsing Hua
 University,\\ Hsinchu, Taiwan R.O.C. \\
E-mail: cheung@phys.cts.nthu.edu.tw}

\maketitle

\abstracts
{
In this talk, I summarize a work done in collaboration \cite{CK} 
with Otto Kong on the
Zee neutrino mass model.  We show that the MSSM with explicit $R$-parity
violation actually contains the Zee model with the right-handed sleptons
as the Zee singlet.  We determine the conditions on the parameter space such
that the neutrino mass matrix provides a viable texture that explains the 
atmospheric and solar data.
}

\section{Introduction}
We have seen substantial amount of experimental evidences from 
solar and atmospheric neutrino experiments
that neutrinos  in fact have masses.
Among the experiments, SuperKamiokande \cite{superk}
provided the strongest evidence for the 
atmospheric neutrino deficit, especially the impressive zenith
angle distribution.  
The neutrino oscillation of $\nu_\mu \to \nu_\tau$ provides
the best explanation for the atmospheric neutrino deficit.  On the other hand,
the solar neutrino deficit is best explained by $\nu_e \to \nu_\mu, \nu_\tau$.

So where do we stand if neutrinos do in fact oscillate?  
\begin{enumerate}
\item Neutrino oscillation necessarily implies neutrinos have masses and of 
different masses.

\item  However, we do not know the absolute values of the masses.  We only 
know the mass differences.  The mass difference required to explain the
atmospheric neutrino deficit is \cite{fits}
\begin{displaymath}
\Delta m_{\rm atm}^2 \sim 3 \cdot 10 ^{-3} \;\;{\rm eV}^2 \;,
\end{displaymath}
while a few solutions to the solar neutrino deficit exist.
For example, the LMA solution requires  a mass difference of \cite{fits}
\begin{displaymath}
\Delta m_{\rm solar}^2 \sim 10 ^{-5} \;\;{\rm eV}^2 \;\;\;\; ({\rm MSW}) \;.
\end{displaymath}

\item Though we do not know the absolute mass scale of the neutrinos,
we have indirect constraints from various sources.  The cosmological 
constraint $\Omega_{\rm hot} \alt 0.1$ implies $m_\nu \alt 3$ eV, assuming
neutrinos make up the hot dark matter.  The end point of Tritium decay also
constrains $m_{\nu_e} \alt 2.2$ eV. Nevertheless, the best constraint comes
from the neutrinoless double beta ($0\nu \beta\beta$)decay.  The absence of 
$0\nu \beta\beta$ decay put an upper bound on the effective neutrino mass, as
\begin{displaymath}
\langle m_{\nu} \rangle_e \equiv \sum_i \;
m_{\nu_i}V_{e_i}^2 \alt 0.2 \;\; {\rm eV} \;.
\end{displaymath}

\end{enumerate}

We know of two widely separated mass scales in neutrinos: 
$\Delta m_{\rm atm}^2$ and $\Delta m_{\rm solar}^2$.  Two possibilities 
of arranging the three neutrino masses exist: (1)
$m_1 \ll m_2 \sim m_3$ or (2) $m_1 \sim m_2 \ll m_3$, assuming 
$m_1 < m_2 < m_3$:

\begin{center}
\includegraphics[height=1.4in,width=4.5in]{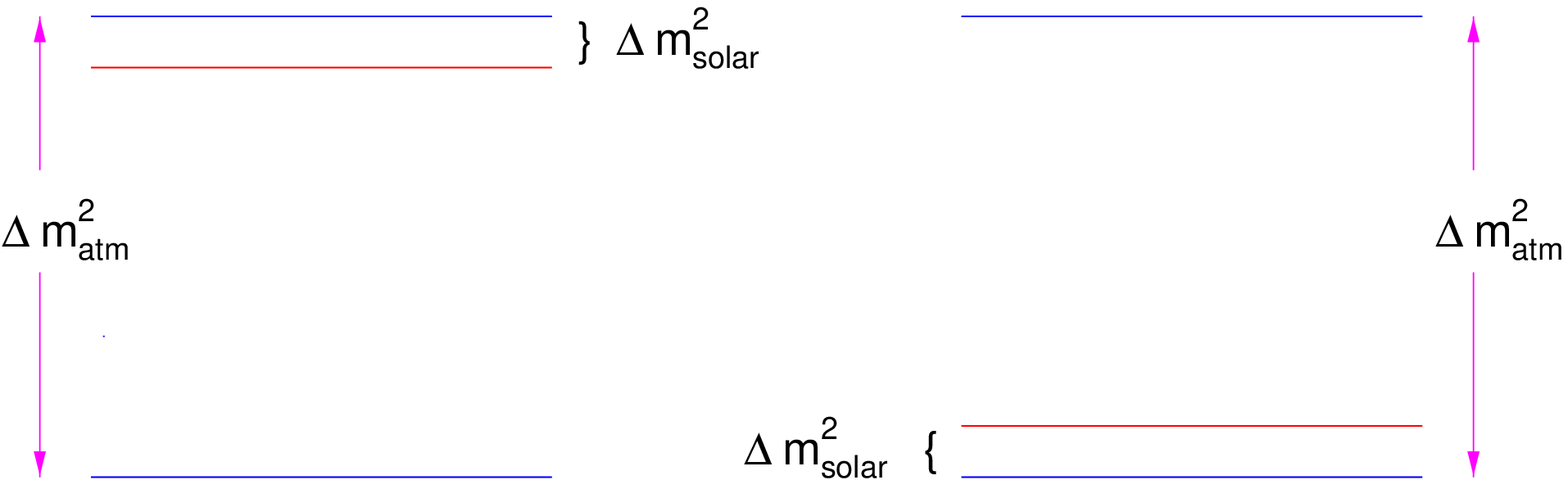}
\end{center}

\section{Types of neutrino mass}

There are three types of neutrino mass according to the structure of
the mass term.

(i) Dirac neutrino mass: $\overline{\psi_L} {\cal M}_D \chi_R +h.c.$,
in which $\chi_R$ is the right-handed neutrino field.  This is 
analogous to the Dirac mass term for charged leptons.  However, this term is
not allowed in the SM, because the bare mass term is forbidden by gauge
invariance and the  SM does not have the right-handed neutrino
field.  Even in the case of charged leptons,  the Dirac mass term must be 
derived from the Yukawa term with a Higgs field or equivalent, in order 
that gauge invariance is fulfilled before the symmetry breaking, followed
by spontaneous symmetry breaking that the Higgs field develops a VEV.

(ii) Left-handed marjorana neutrino mass: $\psi^T_L C^{-1} {\cal M}_L \psi_L$,
where $C$ is the charge conjugation operator.  Again, this bare mass term is
not allowed in the SM due to gauge invariance.  Therefore, it must be derived
from a Yukawa term with a Higgs field or equivalent.  However, in this case
a $I=1, Y=2$ Higgs field is required to generate such a mass term.  SM does
not have such a Higgs field.

(iii) Right-handed marjorana mass: $\chi^T_R C^{-1} {\cal M}_R \chi_L$.
In the SM, there is no right-handed neutrino field.  

Therefore, to generate nonzero neutrino mass one has to include new physics
beyond the SM.  In both (i) and (iii) a right-handed field has to be 
introduced while the case (ii) does not necessarily require a right-handed 
field.  

The hierarchy between the small neutrino mass and the charged lepton
mass tells us something special about the mechanism that generates the neutrino
mass, otherwise a fine tuning of the small Yukawa coupling for neutrinos is
needed.  A natural way to generate small neutrino mass is the see-saw
mechanism, making use of a very large mass scale.   Suppose there
exist heavy right-handed neutrino fields $\chi_R$'s that couple to the 
left-handed neutrino fields via the usual Yukawa coupling.  After electroweak
symmetry breaking,
\begin{equation}
{\cal L} = \overline{\nu_{L_i}} \; \left({\cal M}_D \right)_{ij} \; \chi_{R_j}
+ \chi^T_{R_i} \; C^{-1} \; \left({\cal M}_R \right)_{ij} \; \chi_{R_j} +h.c. 
\;,
\end{equation}
where the first term is the Dirac mass term for the neutrinos and the last
term is the majorana mass for the right-handed fields.  We can then write the
mass matrix as 
\begin{equation}
\frac{1}{2}
( \overline{\nu_L} \;\;\; \overline{\chi^{(c)}_R} ) \; \left (
\begin{array}{cc}
{\cal M}_L=0 \;\;\;{\cal M}_D \\
{\cal M}^T_D \;\;\; {\cal M}_R  \end{array} \right ) \; 
\left ( \begin{array}{c}
 \nu^{(c)}_L \\
 \chi_R \end{array} \right ) \; + h.c.
\end{equation}
After diagonalizing the mass matrix, the mass matrix of the light neutrinos 
is given by
\begin{equation}
M_\nu = - {\cal M}_D {\cal M}_R^{-1} {\cal M}^T_D \;,
\end{equation}
where ${\cal M}_R^{-1}$ is the inverse of the majorana mass matrix.  If
${\cal M}_R$ is sufficiently large, it naturally obtains small neutrino 
mass.   Or equivalently, in terms of a dim-5 operator:
\[
{\cal L} = \frac{y_{ij}}{M_R} (L_i H_2) (L_j H_2) \;.
\]
To explain the observed neutrino mass the scale of ${\cal M}_R \sim 10^{10-13}$
GeV for a typical Yukawa coupling.  Such an intermediate scale arouses a lot
of theoretical speculations and interests.  Should the $\chi_R$ related to
SUSY breaking or early unification (a prediction of the Type I string theory
is that the string scale is around $10^{11}$ GeV.)  

Another natural way to generate small neutrino masses is to make use of 
loop suppression.   This need not introduce right-handed neutrino fields,
though new physics is still needed to generate the neutrino mass.  One nice
example is the Zee model \cite{zee}. 

\section{Zee mass model}

\begin{figure}[t]
\centering
\includegraphics[width=4in]{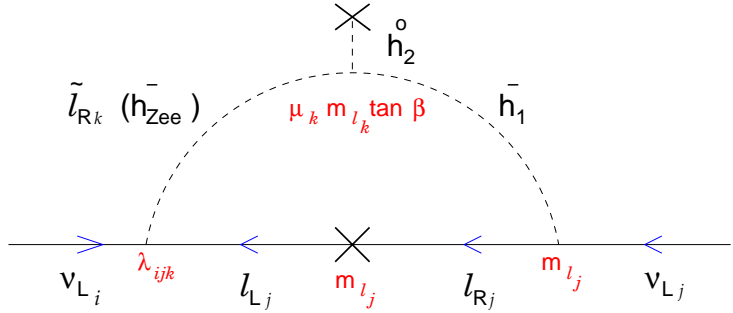}
\caption{\label{fig1} A Feynman diagram for the Zee model, embedded in the 
RPV SUSY framework.}
\end{figure}

Zee model \cite{zee}
provides an economical way to generate small neutrino masses with a 
favorable texture \cite{zee,fg,jmst}.
The model consists of a charged gauge singlet scalar $h^{\mbox{-}}$, 
which couples to lepton doublets 
$\psi_{\!\scriptscriptstyle Lj}$ via the interaction
\begin{equation} \label{zcp}
f^{ij} \left( \psi^\alpha_{\!\scriptscriptstyle Li} {\cal C} 
\psi^\beta_{\!\scriptscriptstyle Lj} \right ) 
\epsilon_{\!\scriptscriptstyle \alpha\beta}\; 
h^{\scriptscriptstyle -} \;,
\end{equation}
where $\alpha,\beta$ are the SU(2) indices, $i,j$ are the generation indices, 
${\cal C}$ is the charge-conjugation matrix, 
and $f^{ij}$ are Yukawa couplings antisymmetric in $i$ and $j$.  
Another ingredient of the model is an extra Higgs doublet (in addition to 
the one that gives masses to charged leptons) that develops a 
VEV and thus provides mixing between
the charged Higgs boson and the Zee singlet. 
The one-loop diagram for the Zee model is depcited  in Fig.~\ref{fig1}.

The Zee model can provide a mass matrix of the following texture \cite{fg,jmst}
\begin{equation} 
\label{zee-m}
 \left( \begin{array}{ccc}
0        & m_{e\mu} & m_{e\tau} \\
m_{e\mu} & 0        & \epsilon  \\
m_{e\tau}& \epsilon & 0 
         \end{array}          \right ) \; ,
\end{equation}
where $\epsilon$ is small compared with $m_{e\mu}$ and $m_{e\tau}$, which is
able to provide a compatible mass pattern that explains the atmospheric
and solar neutrino data. Diagonal elements are guaranteed to vanish
while the $m_{\mu\tau}$ entry, denoted by $\epsilon$, has to be 
suppressed by some means.  Moreover, $m_{e\mu} \sim m_{e\tau}$ is required to
give the maximal mixing solution for the atmospheric neutrinos.

First, take $\epsilon=0$ the matrix can be diagonalized by 
\begin{equation}
\left( \begin{array}{c}
         \nu_{\!\scriptscriptstyle Le} \\
         \nu_{\!\scriptscriptstyle L\mu} \\
         \nu_{\!\scriptscriptstyle L\tau} \end{array} \right )
=
\left( \begin{array}{ccc}
   \frac{1}{\sqrt{2}} &  \frac{1}{\sqrt{2}} &  0 \\
   \frac{m_{e\mu}}{\sqrt{2} m} & \frac{-m_{e\mu}}{\sqrt{2} m} &
   \frac{- m_{e\tau}}{m}  \\
   \frac{m_{e\tau}}{\sqrt{2} m} & \frac{-m_{e\tau}}{\sqrt{2} m} &
   \frac{m_{e\mu}}{m}  \end{array} \right ) \;
\left( \begin{array}{c}
         \nu_{\!\scriptscriptstyle L_1} \\
         \nu_{\!\scriptscriptstyle L_2} \\
         \nu_{\!\scriptscriptstyle L_3} \end{array} \right ) \;,
\end{equation}
with the eigenvalues $m, -m, 0$ for $\nu_{\!\scriptscriptstyle L_1},
\nu_{\!\scriptscriptstyle L_2},\nu_{\!\scriptscriptstyle L_3}$, respectively,
and $m=\sqrt{m_{e\mu}^2 + m_{e\tau}^2}$. The atmospheric mass-squared 
difference
$\Delta m^2_{\rm atm} \simeq  3\times 10^{-3} {\rm eV}^2$, is to be 
identified with 
$m^2=m_{e\mu}^2 + m_{e\tau}^2$.  The transition probabilities for 
$\nu_{\!\scriptscriptstyle L_\mu}$ are
\begin{eqnarray}
P_{\nu_{\scriptscriptstyle L_\mu} \to \nu_{\scriptscriptstyle L_e} } &=& 0 
\;, \nonumber \\
P_{\nu_{\scriptscriptstyle L_\mu} \to \nu_{\scriptscriptstyle L_\tau} } 
&=& 4 \left( \frac{m_{e\mu} m_{e\tau} }
{ m_{e\mu}^2 + m_{e\tau}^2 } \right )^2 \sin^2  \left(
\frac{ (m_{e\mu}^2 + m_{e\tau}^2) L }{4 E} \right )  \;. \nonumber 
\end{eqnarray}
If $m_{e\mu}\simeq m_{e\tau}$, then $\sin^2 2\theta_{\rm atm} \simeq 1$.  This
provides the maximal mixing solution for the atmospheric neutrino anomaly.

If we choose a nonzero $\epsilon$, but keep $\epsilon \ll m_{e\mu, e\tau}$.
Then after diagonalizing the matrix we have the following eigenvalues
\begin{eqnarray}
m_{\nu\scriptscriptstyle 1} &=& \sqrt{m_{e\mu}^2 + m_{e\tau}^2} + 
  \epsilon \frac{m_{e\mu} m_{e\tau}}{ m_{e\mu}^2 + m_{e\tau}^2 }\;, \nonumber\\
m_{\nu\scriptscriptstyle 2} &=& -\sqrt{m_{e\mu}^2 + m_{e\tau}^2} + 
  \epsilon \frac{m_{e\mu} m_{e\tau}}{ m_{e\mu}^2 + m_{e\tau}^2 }\;, \nonumber\\
m_{\nu\scriptscriptstyle 3} &=& 
-2 \epsilon \frac{m_{e\mu} m_{e\tau}}{ m_{e\mu}^2 + m_{e\tau}^2 } 
\;.\nonumber
\end{eqnarray}
The mass-square difference between $m^2_{\nu\scriptscriptstyle 1}$ 
and $m^2_{\nu\scriptscriptstyle 2}$ can be fitted to the
solar neutrino mass.  If one takes the LMA solution and requires
\begin{displaymath}
4 \epsilon \frac{m_{e\mu} m_{e\tau}}{\sqrt{ m_{e\mu}^2 + m_{e\tau}^2 }} =
 \Delta m^2_{\rm sol} \simeq 2 \times 10^{-5} \;{\rm eV}^2 \; ,
\end{displaymath}
giving (we have used $m_{e\mu} \simeq m_{e\tau}$)
\begin{displaymath}
\frac{\epsilon}{m_{e\mu}} \sim 5\times 10^{-3} \;.
\end{displaymath}

\section{Neutrino mass in SUSY}

The original Zee model was not embedded into any grand unified theories or
supersymmetric models.  It would be very interesting if the Zee model 
naturally exists in some GUT  or SUSY theories.  In fact, the 
minimal supersymmetric standard model (MSSM) with a minimal 
extension, namely, the $R$-parity violation, contains the Zee model.
  The right-handed sleptons 
in SUSY have the right quantum numbers to play the role of the charged 
Zee singlet.  The $R$-parity-violating (RPV) $\lambda$-type couplings
could provide the terms in Eq.(\ref{zcp}). It is also easy to see that 
the RPV bilinear $\mu$-type couplings ($\mu_i 
L H_{\scriptscriptstyle 2}$) 
would allow the second Higgs doublet $H_{\scriptscriptstyle 2}$ in
SUSY to be the second ingredient of the Zee model.

However, in RPV SUSY framework, there are three 
other sources for neutrino masses, 
in addition to the Zee model contribution.  
They are
(i) the tree-level mixing with the higgsinos and gauginos, (ii) the 
one-loop diagram that involves the usual mass mixing between the left-handed 
and right-handed sleptons proportional to $m_{\scriptscriptstyle \ell}\, 
(A^{\!\scriptscriptstyle E}_{\scriptscriptstyle \ell} - \mu \tan\beta)$,
and (iii) the one-loop diagram that again involves the mixing between 
the left-handed and right-handed sleptons but this time via the 
$\lambda$ and $\mu_i$ couplings.
They may deviate from the texture of the Zee mass matrix of Eq. (\ref{zee-m}). 

{\it The tree-level mixing}
 among the higgsinos, gauginos, and neutrinos gives rise
to a $7\times 7$ neutral fermion mass matrix $\cal{M_N}$ under SVP \cite{svp}:
\begin{equation}
\label{77}
\cal{M_N} = 
\left (\begin{array}{cccc|ccc}
M_{\scriptscriptstyle 1} & 0 & g'v_{\scriptscriptstyle 2}/2 
& -g' v_{\scriptscriptstyle 1}/2 & 0 & 0 & 0  \\
0   & M_{\scriptscriptstyle 2} & -gv_{\scriptscriptstyle 2}/2 
& g v_{\scriptscriptstyle 1}/2 & 0 & 0 & 0  \\
g'v_{\scriptscriptstyle 2}/2 & -g v_{\scriptscriptstyle 2}/2 & 0   
& -\mu & -\mu_{\scriptscriptstyle 1} & -\mu_{\scriptscriptstyle 2} 
& -\mu_{\scriptscriptstyle 3} \\
-g'v_{\scriptscriptstyle 1}/2 & g v_{\scriptscriptstyle 1}/2 & -\mu 
&   0 & 0 & 0 & 0\\
\hline
0 & 0 & -\mu_1     & 0 & (m_\nu^0)_{\!\scriptscriptstyle 1\!1} 
& (m_\nu^0)_{\!\scriptscriptstyle 1\!2} & (m_\nu^0)_{\!\scriptscriptstyle 1\!3}
\\
0 & 0 &  -\mu_{\!\scriptscriptstyle 2} & 0  & (m_\nu^0)_{\!\scriptscriptstyle 
2\!1} 
& (m_\nu^0)_{\!\scriptscriptstyle 22} & (m_\nu^0)_{\!\scriptscriptstyle 23} \\
0 & 0 & -\mu_{\!\scriptscriptstyle 3} & 0 & (m_\nu^0)_{\!\scriptscriptstyle 31}
& (m_\nu^0)_{\!\scriptscriptstyle 32} & (m_\nu^0)_{\!\scriptscriptstyle 33} \\
 \end{array}
\right )\; ,
\end{equation}
whose basis is $(-i\tilde{B}, -i\tilde{W}, 
\tilde{h}_{\scriptscriptstyle 2}^0, \tilde{h}_{\scriptscriptstyle 1}^0, 
\nu_{\!\scriptscriptstyle L_e},\nu_{\!\scriptscriptstyle L_\mu },
\nu_{\!\scriptscriptstyle  L_\tau}) $. 

In the above $7\times 7$ 
matrix, the whole lower-right $3\times 3$ block $(m_\nu^0)$ is zero at 
tree level.  They are induced via one-loop contributions.
We can write the mass matrix in the form of block submatrices:
\begin{equation}
{\cal M_N} = \left( \begin{array}{c|c}
              {\cal M} & \xi^{\!\scriptscriptstyle  T} \\
\hline
              \xi & m_\nu^0 \end{array}  \right ) \;,
\end{equation}
where $\cal{M}$ is the upper-left $4\times 4$ neutralino mass matrix, 
$\xi$ is the $3\times 4$ block, and $m_\nu^0$ is the lower-right 
$3\times 3$ neutrino block in the $7\times 7$ matrix.  
The resulting neutrino mass matrix after block diagonalization is given by
\begin{equation} \label{mnu}
(m_\nu) = - \xi {\cal M}^{\mbox{-}1} \xi^{\!\scriptscriptstyle  T} + (m_\nu^0)
 \;.
\end{equation}
The first term here corresponds to the tree level contributions, which are 
see-saw suppressed.

Through this gaugino-higgsino mixing, 
nonzero $\mu_i$'s give tree-level see-saw type contributions to 
$(m_\nu)_{ij}$ proportional to  $\mu_i \mu_j$, given by
\begin{equation} \label{mnu2}
(m_\nu)_{ij} =  - \,\frac{v^2 \cos^2\!\beta 
\;( g^2 M_{\scriptscriptstyle 1} + g^{'2} M_{\scriptscriptstyle 2} )}
{2 \mu \;[2 \mu M_{\scriptscriptstyle 1} M_{\scriptscriptstyle 2} 
- v^2 \sin\!\beta \cos\!\beta \;(g^2 M_{\scriptscriptstyle 1} 
+ g^{'2}M_{\scriptscriptstyle 2} ) ]}\;\mu_i \mu_j \;.
\end{equation}
A diagonal $(m_\nu)_{kk}$ term is 
always present for a nonzero $\mu_k$.
To eliminate these tree-level terms requires either very stringent constraints
on the parameter space or extra Higgs superfields 
beyond the MSSM spectrum. This is a major difficulty of 
the present MSSM formulation of supersymmetric Zee model.

{\it Zee mechanism}. 
The Feynman diagram is shown in Fig. \ref{fig1}.
The $\tilde{\ell}_{\!\scriptscriptstyle R_k}$
is the charged Zee singlet.
To complete the diagram the 
charged Higgs boson $h_{\scriptscriptstyle 1}^{\mbox{-}}$ from the Higgs 
doublet $H_{\scriptscriptstyle 1}$ is on 
the other side of the loop and a 
$\tilde{\ell}_{\!\scriptscriptstyle R_k}$-$h_1^{\mbox{-}}$ mixing
at the top of the loop is provided by a  $F$ term of $L_k$:
$\mu_k m_{\!\scriptscriptstyle \ell_k} h_{\scriptscriptstyle 1}^{\mbox{-}} 
\tilde{\ell}^*_{\!\scriptscriptstyle R_k} \langle h_{\scriptscriptstyle 2}^0
\rangle / \langle h_{\scriptscriptstyle 1}^0 \rangle$,
where $h_{\scriptscriptstyle 2}^0$ takes on its VEV, for a nonzero 
$\mu_k$.  Thus, the neutrino mass term $(m_\nu^0)_{ij}$ has a
\begin{equation}
\mu_k m_{\!\scriptscriptstyle \ell_{k}}\lambda_{ijk}
(m_{\!\scriptscriptstyle \ell_j}^2 -m_{\!\scriptscriptstyle \ell_i}^2)
\end{equation}
dependence, where $m_{\!\scriptscriptstyle \ell_i}$'s
are the charged lepton masses. 

{\it $LR$ slepton mass mixing}
comes from the one-loop diagram 
with two $\lambda$-coupling vertices and the usual
$(A^{\!\scriptscriptstyle E}-\mu\tan\beta)$-type 
$LR$ slepton mixing. Neglecting the
off-diagonal entries in $A^{\!\scriptscriptstyle E}$, 
the contribution to $(m_\nu^0)_{ij}$ with the pair $\lambda_{ilk}$ and 
$\lambda_{jkl}$ is proportional to 
\begin{equation}
\label{Aloop}
 \big[ \;(A^{\!\scriptscriptstyle E}_k - \mu \tan\!{\beta})
+ (1-\delta_{kl}) (A^{\!\scriptscriptstyle E}_l - \mu \tan\!{\beta})\;\big]\;
m_{\!\scriptscriptstyle \ell_k} m_{\!\scriptscriptstyle \ell_l} 
 \lambda_{ilk}\lambda_{jkl}  \;.
\end{equation}

{\it $LR$ slepton mass mixing via RPV couplings}
comes from a  $F$ term of $L_i$: $\mu_i\lambda_{ijk} 
\tilde{\ell}_{\!\scriptscriptstyle L_j} \tilde{\ell}^*_{\!\scriptscriptstyle 
R_k} 
\langle h_{\scriptscriptstyle 2}^0 \rangle$, where 
$h_{\scriptscriptstyle 2}^0$ takes on the VEV.  
This is similar to the $\tilde{\ell}_{\!\scriptscriptstyle R_k}
\!\mbox{-}h_{\scriptscriptstyle 1}^{\mbox{-}}$ mixing in the Zee model,
except that this time we have a $\lambda$-type coupling in place of
the Yukawa coupling.  
With a specific choice of a set of nonzero 
$\mu_i$'s and $\lambda$'s, this type of mixing gives rise to 
the off-diagonal $(m_\nu^0)_{ij}$ terms only and, therefore,
of particular interest to our perspectives of Zee model.
Taking the pair $\lambda_{ilk}$ and $\lambda_{jhl}$ for the fermion
vertices and a  
$F$ term of $L_g$ providing a coupling for the scalar vertex
in the presence of a $\mu_g$ and a $\lambda_{ghk}$,
a $(m_\nu^0)_{ij}$ term is generated and proportional to 
\begin{equation}
\label{ghk}
\mu_g m_{\!\scriptscriptstyle \ell_l}\lambda_{ghk} \lambda_{ilk} \lambda_{jhl}
 \; .
\end{equation}
When we allow only one nonzero $\lambda$ at a time, the only 
contribution comes from $\lambda_{ijj}$ but not from those with distinct 
indices.  Suppose we have nonzero $\lambda_{ijj}$  and $\mu_j$, 
there is a contribution to the off-diagonal $(m_\nu^0)_{ij}$ with a
$\mu_j m_{\!\scriptscriptstyle \ell_j} \lambda_{ijj}^3$
dependence.

We conclude that a minimal set of RPV
 couplings needed to give the zeroth order Zee matrix is
\[
\{  \quad
\lambda_{{\scriptscriptstyle 12}\,k}\;,\quad
\lambda_{{\scriptscriptstyle 13}\,k}\;,\quad
\mu_k \quad \}\; .
\]
As at least one of the two $\lambda$'s has the form $\lambda_{ikk}$
($\equiv -\lambda_{kik}$), 
all types of contributions that have been discussed above are there. 
We want to make the contribution from the Zee mechanism dominate over the 
others, or at least to suppress the diagonal entries in $(m_\nu)$.
This necessarily requires suppression of the 
contributions from the tree-level see-saw mechanism and from the 
$(A^{\!\scriptscriptstyle E} - \mu \tan\!{\beta})$-type $LR$ slepton mixing. 
So, it is the Zee mechanism and the $LR$ mixing via RPV
couplings are required to be the dominant ones.

\section{Scenarios and conditions to maintain Zee Texture}

Because of space limitation we only show the best scenario:
$\{\lambda_{\scriptscriptstyle 1\!23}$, 
$\lambda_{\scriptscriptstyle 1\!33}$,  and $\mu_{\scriptscriptstyle 3}\}$.
The resulting neutrino mass matrix is given by
\small \begin{equation}
\label{scen2}
\left( \begin{array}{ccc}
C'_{\scriptscriptstyle 4}\, m_\tau^2\, \lambda_{\scriptscriptstyle 1\!33}^2
\;\; & \;\;
C'_{\scriptscriptstyle 2} \, m_\tau \, m_\mu^2 \,
\mu_{\scriptscriptstyle 3} \lambda_{\scriptscriptstyle 1\!23} 
+ C_{\scriptscriptstyle 5}\;
 m_\tau \, \mu_{\scriptscriptstyle 3} \lambda_{\scriptscriptstyle 1\!23}
 \lambda_{\scriptscriptstyle 1\!33}^2
\;\; & \;\;
   C'_{\scriptscriptstyle 2} \, m_\tau^3 \, 
\mu_{\scriptscriptstyle 3} \lambda_{\scriptscriptstyle 1\!33} 
+ C_{\scriptscriptstyle 5}\;
 m_\tau \, \mu_{\scriptscriptstyle 3}  \lambda_{\scriptscriptstyle 1\!33}^3
 \\    & 0 & 0 \\
   & & C_{\scriptscriptstyle 1} \;\mu_{\scriptscriptstyle 3}^2  
\end{array} \right )\;
\end{equation} \normalsize
where
\begin{eqnarray}
C_{\scriptscriptstyle 1} &=& - \,\frac{v^2 \cos^2\!\beta \;
( g^2 M_{\scriptscriptstyle 1} + g^{'2} M_{\scriptscriptstyle 2} )}
{2 \mu \;[2 \mu M_{\scriptscriptstyle 1} M_{\scriptscriptstyle 2} 
- v^2 \sin\!\beta \cos\!\beta\; (g^2 M_{\scriptscriptstyle 1} 
+ g^{'2}M_{\scriptscriptstyle 2} ) ]} \;, \nonumber \\
C'_{\scriptscriptstyle 4} &=& - \frac{1}{16\pi^2} \; 
( A^{\!\scriptscriptstyle E}_\tau - \mu \tan\!\beta) 
\;f(M^2_{\tilde{\tau}_L},M^2_{\tilde{\tau}_R}) 
 \;, \nonumber \\
 C'_{\scriptscriptstyle 2} &=&\frac{-1}{16\pi^2} 
\frac{\sqrt{2}\tan\!\beta} {v \cos\!\beta}\;  
f( M_{h_{\scriptscriptstyle 1}^{\mbox{-}}}^2, M_{\tilde{\tau}_R}^2 )
\; , \nonumber \\
C_{\scriptscriptstyle 5} &=&- \frac{1}{16\pi^2}  \frac{v\sin\!\beta}{\sqrt{2}}
 f( M_{\tilde{e}_L}^2, M_{\tilde{\tau}_R}^2 )\;,
\end{eqnarray} 
where
$f(x,y) = \frac{1}{x-y} \; \log ( y/x ) $.

In the above, we have neglected terms suppressed by $m_e/m_\mu$ or 
$m_e/m_\tau$.  
In order to maintain the zeroth order Zee texture,
we need $m_{e\mu}$ and $m_{e\tau}$ to dominate over the other
entries. Moreover, we need $m_{e\mu}\sim m_{e\tau}\sim\sqrt{\Delta
M^2_{\rm atm}} (\sim 5\times 10^{-11}\;\mbox{GeV})$.  

Requiring the tree-level gaugino-higgsino mixing contribution 
to be well below $m_{e\mu}$ gives
\begin{equation} \label{mtt}
{\mu_{\scriptscriptstyle 3}^2}\, \cos^2\!\!\beta \ll 
{\mu^2} M_{\scriptscriptstyle 1}\,(1\times 10^{-14}\, {\rm GeV}^{-1}) \;.
\end{equation} 

For the $(A^{\scriptscriptstyle E}_k - \mu \tan\!\beta)$ $LR$ slepton mixing
contribution to be much smaller than $m_{e\mu}$, we have 
\begin{equation} \label{2a}
\lambda_{\scriptscriptstyle 1\!33}^2
\ll \frac{\mbox{Max}(M^2_{\tilde{\tau}_L},M^2_{\tilde{\tau}_R})}
{(A^{\scriptscriptstyle E}_\tau - \mu \tan\!\beta)}\;
 (2.5 \times 10^{-9} \,\mbox{GeV}^{-1})\;.
\end{equation}
This corresponds to $m_{ee}$. 
It tells us that $\lambda_{\scriptscriptstyle 1\!33}$ can hardly be much 
larger than $10^{-3}$.
On the other hand, $\lambda_{\scriptscriptstyle 1\!23}$ is constrained
differently because  
it does not contribute to this type of neutrino mass term.

{}From the tree-level Zee-scalar mediated $\mu$ decay,
the constraint is 
\begin{equation} \label{mudy1}
\frac{\lambda_{\scriptscriptstyle 1\!23}^2}{M^2_{\tilde{\tau}_R}}
\leq 10^{-8}\,\mbox{GeV}^{-2} \;,
\end{equation}
which tells us that $\lambda_{\scriptscriptstyle 1\!23}$ can be as large as
order of $0.01$ for scalar masses of order of $O(100)$ GeV. 

 Both $m_{e\mu}$ and $m_{e\tau}$ have two terms.  Let us look at
$m_{e\mu}$ first.
For the first term in $m_{e\mu}$ (the one with a $C'_{\scriptscriptstyle 2}$
dependence) in Eq. (\ref{scen2}) to give the required value of
atmospheric neutrino mass, we need 
\begin{equation} \label{2zm12}
m_{e\mu}\;\sim\;
\frac{\mu_{\scriptscriptstyle 3} \lambda_{\scriptscriptstyle 1\!23}}
{\cos^2\!\!\beta}\;
\frac{1}{\mbox{max}( M_{h_{\scriptscriptstyle 1}^{\mbox{-}}}^2, 
M_{\tilde{\tau}_R}^2 )} \;
(7\times 10^{-7}\,\mbox{GeV}^{2})\; 
\sim \;(5\times 10^{-11}\, {\rm GeV})
\end{equation}
or
\begin{equation}
(\mu_{\scriptscriptstyle 3} \cos\!\beta) \;\lambda_{\scriptscriptstyle 1\!23}
\sim \;{\cos^3\!\!\beta}\;{\mbox{max}( 
M_{h_{\scriptscriptstyle 1}^{\mbox{-}}}^2, 
M_{\tilde{\tau}_R}^2 )}\; (7\times 10^{-5} \,\mbox{GeV}^{-1})\;.
\end{equation}
This result looks relatively promising. If we take $\cos\!\beta=0.02$,
all the involved scalar masses at $100\,\mbox{GeV}$ and 
$\lambda_{\scriptscriptstyle 1\!23}$ at the corresponding limiting $0.01$ 
value, $\mu_{\scriptscriptstyle 3} \cos\!\beta$ has to be at $5.6\times 10^{-4}
\,\mbox{GeV}$ to fit the requirement. This means pushing for larger
$M_{\scriptscriptstyle 1}$ (and  $M_{\scriptscriptstyle 2}$) and $\mu$ values
but may not be ruled out. 

The corresponding first term in $m_{e\tau}$ has a 
$\lambda_{\scriptscriptstyle 1\!33}$ dependence in the place of 
$\lambda_{\scriptscriptstyle 1\!23}$ with  an extra enhancement of 
$m_\tau^2/m_\mu^2$, in comparison to $m_{e\mu}$. That is to say, requiring
$m_{e\mu}\approx m_{e\tau}$ gives, in this case,
\begin{equation}
\label{29}
\lambda_{\scriptscriptstyle 1\!33} \approx \frac{m_\mu^2}{m_\tau^2}\;
\lambda_{\scriptscriptstyle 1\!23} \; .
\end{equation}
This gives a small $\lambda_{\scriptscriptstyle 1\!33}$ easily satisfying
Eq. (\ref{2a}). The small $\lambda_{\scriptscriptstyle 1\!33}$ also suppresses
the second terms in both $m_{e\mu}$ and $m_{e\tau}$, the
$C_{\scriptscriptstyle 5}$ dependent terms in Eq. (\ref{scen2}).

To produce the neutrino mass matrix beyond the zeroth order Zee texture, 
the subdominating first-order contributions are required to be 
substantially smaller in order to fit the solar neutrino data. Here, 
it is obvious that it is difficult to further suppress the tree level 
gaugino-higgsino mixing contribution to $m_{\tau\tau}$, which makes it 
even more difficult to get the scenario to work. Explicitly, the requirement
for the solar neutrino is
\begin{equation} 
{\mu_{\scriptscriptstyle 3}^2}\, \cos^2\!\!\beta \sim 
{\mu^2} M_{\scriptscriptstyle 1}\,(1\times 10^{-16}\, {\rm GeV}^{-1}) \;.
\end{equation} 

\section{A general version of supersymmetric Zee model}
The conditions for maintaining
the Zee neutrino mass matrix texture is extremely stringent, if not impossible,
mainly because of the tree-level mixings via the bilinear RPV couplings.
An alternative without the bilinear RPV
couplings is to introduce an additional pair of Higgs doublet superfields. 
Denoting 
them by $H_{\!\scriptscriptstyle 3}$ and $H_{\!\scriptscriptstyle 4}$, 
bearing the same quantum numbers as $H_{\!\scriptscriptstyle 1}$ and 
$H_{\!\scriptscriptstyle 2}$, respectively, RPV terms
of the form
\[
\epsilon_{\!\scriptscriptstyle \alpha\beta} \lambda^{\!\scriptscriptstyle H}_k
H_{\!\scriptscriptstyle 1}^\alpha H_{\!\scriptscriptstyle 3}^\beta
E_k^c 
\]
can be introduced. With a trivial extension of notations we obtain 
a Zee diagram contribution to $(m_\nu)_{ij}$ through $\lambda_{ijk}$ as 
follows :
\begin{equation} \label{h1i}
\frac{-1}{16\pi^2} \, \frac{\langle h_{\scriptscriptstyle 3}^0 \rangle} 
{\langle h_{\scriptscriptstyle 1}^0 \rangle}\;  
(m_{\scriptscriptstyle \ell_j}^2 -m_{\scriptscriptstyle \ell_i}^2) \;
\lambda_{ijk}\;
\lambda^{\!\scriptscriptstyle H}_k A^{\!\scriptscriptstyle H}_k  \;
f( M_{h_{\scriptscriptstyle 1}^{\mbox{-}}}^2, M_{\tilde{\ell}_{R_k}}^2 ) \;.
\end{equation}
Here the slepton $\tilde{\ell}_{\!\scriptscriptstyle R_k}$ keeps the role
of the Zee singlet.  Notice that
the second Higgs doublet of the Zee model, corresponding to 
$H_{\!\scriptscriptstyle 3}$ here, is assumed not to have couplings
of the form $L_i H_{\!\scriptscriptstyle 3} E^c_j$. The condition for the
$LR$ slepton mixing contribution to be below the required $m_{e\mu}$
would be the same as discussed in the last section. 

However, there is a new contribution to $(m_\nu)_{kk}$ given by
\begin{equation} \label{hkk}
\frac{-1}{16\pi^2} \, \frac{\langle h_{\scriptscriptstyle 3}^0 \rangle\!^2} 
{ \langle h_{\scriptscriptstyle 1}^0 \rangle\!^2}\; 
m_{\!\scriptscriptstyle \ell_k}^2
(\lambda^{\!\scriptscriptstyle H}_k)^2 A^{\!\scriptscriptstyle H}_k  \;
f( M_{h_{\scriptscriptstyle 1}^{\mbox{-}}}^2, M_{\tilde{\ell}_{R_k}}^2 ) \;.
\end{equation}
This is a consequence of the fact that the 
term $\lambda^{\!\scriptscriptstyle H}_k
H_{\!\scriptscriptstyle 1}^\alpha H_{\!\scriptscriptstyle 3}^\beta
E_k^c$ provides new mass mixings for the charged 
Higgsinos and the charged leptons.
The essential difference here is that unlike the $\mu_i$ terms the 
$\lambda^{\!\scriptscriptstyle H}_k 
H_{\!\scriptscriptstyle 1}^\alpha H_{\!\scriptscriptstyle 3}^\beta E_k^c$ 
term does not contribute to the
mixings between neutrinos and the gauginos and higgsinos on tree level.

Similar to the above
we are interested in only the minimal set of couplings 
$\{\lambda_{{\scriptscriptstyle 12}\,k}\;,\;
\lambda_{{\scriptscriptstyle 13}\,k}\;,\; 
\lambda^{\!\scriptscriptstyle H}_{\scriptscriptstyle k}\}$ with a specific $k$.
For expression (\ref{h1i}) to give the right value to $m_{e\mu}$, we need
\begin{equation}
\lambda_{{\scriptscriptstyle 12}\,k}\, \lambda^{\!\scriptscriptstyle H}_k
\; \sim \frac{\mbox{Max}( M_{h_{\scriptscriptstyle 1}^{\mbox{-}}}^2, 
M_{\tilde{\ell}_{R_k}}^2 )}{A^{\!\scriptscriptstyle H}_k}\;
\frac{\langle h_{\scriptscriptstyle 1}^0 \rangle}
{\langle h_{\scriptscriptstyle 3}^0 \rangle} 
\; (7\times 10^{-7} \,\mbox{GeV}^{-1})\;, 
\end{equation}
and similarly for $m_{e\tau}$, it requires
$\lambda_{{\scriptscriptstyle 13}\,k}= (m_\mu^2/m_\tau^2)
\lambda_{{\scriptscriptstyle 12}\,k}\;$. This condition is easy to satisfy
when we take  $\langle h_{\scriptscriptstyle 3}^0 \rangle /
 \langle h_{\scriptscriptstyle 1}^0 \rangle  = 0.1$. 
For Eq. (\ref{h1i}) to dominate over Eq. (\ref{hkk}), it requires
\begin{equation}
\lambda_{{\scriptscriptstyle 12}\,k} \gg \lambda^{\!\scriptscriptstyle H}_k
\; \frac{\langle h_{\scriptscriptstyle 3}^0 \rangle} 
{\langle h_{\scriptscriptstyle 1}^0 \rangle}\; 
\frac{m^2_{\scriptscriptstyle \ell_k}}{m^2_\mu} \;,
\qquad
\lambda_{{\scriptscriptstyle 13}\,k} \gg \lambda^{\!\scriptscriptstyle H}_k
\; \frac{\langle h_{\scriptscriptstyle 3}^0 \rangle} 
{\langle h_{\scriptscriptstyle 1}^0 \rangle}\;
 \frac{m^2_{\scriptscriptstyle \ell_k}}{m^2_\tau} \;.
\end{equation}
The most favorable scenario is then the $k=1$ case,
where $m_{\scriptscriptstyle \ell_k}$ is just the $m_e$. The above requirements
are then easily satisfied.  Also, the requirement for
suppression of the $LR$ slepton mixing is the same as before,
and we also have Eq. (\ref{mudy1}) from the tree-level Zee-scalar induced 
muon decay.  All these constraints
can now be easily satisfied. Hence, such a supersymmetric Zee model
looks very feasible.

\section{Conclusions}

Zee model provides a viable texture that explains the data.
The minimal extension of MSSM with $R$-parity violation
actually contains the Zee model, with
the right-handed sleptons $\tilde{\ell}_R$ as the charged singlet,
$\lambda_{ijk}$ couplings providing lepton-number violation, and 
$H_u$ providing the mixing.

However, there are other sources of neutrino mass in RPV
SUSY, some of which wipe away the favorable texture.  In order for the
Zee contribution to dominate over the others we pick the best minimal scheme 
$\{\lambda_{{\scriptscriptstyle
12}\,k}\;,\; \lambda_{{\scriptscriptstyle 13}\,k}\;,\;
\mu_{\scriptscriptstyle k}\}$, $k=3$, 
and determine the requirements on the parameter space, which turns out
quite stringent but still possible.

Finally, we offered a further consideration that abandons the bilinear 
RPV couplings but introduces two additional Higgs doublets.  This model turns 
out quite feasible.

I would like to thank Otto Kong for the pleasant collaboration on the work
presented here.

\end{document}